\documentstyle[12pt,eqsecnum,multicol,psfig,aps,prl,array]{revtex}
\newcommand{\bleq}{\ifpreprintsty
		   \else
		   \end{multicols}\widetext \vspace*{-3.5ex}{\tiny
		   
		\noindent\begin{tabular}[t]{c|}
		   \parbox{0.493\hsize}{~} \\ \hline \end{tabular}}
				      \fi}
\newcommand{\eleq}{\ifpreprintsty
		   \else
		   {\tiny\hspace*{\fill}\begin{tabular}[t]{|c}\hline
		    \parbox{0.49\hsize}{~} \\
		    \end{tabular}}\vspace*{-2.5ex}\begin{multicols}{2}
		    \narrowtext
		    \fi}
\newcommand{\bcols}{\ifpreprintsty\else\begin{multicols}{2} 
	\narrowtext\fi}
\newcommand{\ecols}{\ifpreprintsty\else\end{multicols}\fi}

\draft
\widetext
\begin{document}
\title{Simple rules for the prediction of the value of the glass transition temperature in network 
glasses} 
\author{Matthieu Micoulaut $^{\S}$ and Gerardo G. Naumis $^{\P}$}
\address{\ \\
$^{\S}$ Laboratoire GCR-CNRS URA 769\\
Universit\'e Pierre et Marie Curie, Tour 22, Boite 142\\
4, Place Jussieu, 75252 Paris Cedex 05, France\\
$^{\P}$ Instituto de Fisica, Universidad Nacional Autonoma de M\'exico\\
Apartado Postal 20-364, 01000 M\'exico DF, M\'exico}

\date{\today}
\maketitle
\begin{abstract}
We give in this letter a set of general rules which allow the prediction
of the value of the glass transition temperature $T_g$ in network glasses.
Starting form the Gibbs-Di Marzio law which gives a very general relationship 
between this temperature and the average coordination number of a system, we 
explain how to compute from the valencies of the 
atoms of the glass, the parameter $\beta$ used in this law. We check the validity of 
the obtained expression and show that it is possible to predict the
glass transition temperature for any composition in multicomponent 
chalcogenide glasses. The possibility of existence of a demixed structure is
also discussed.
\par
Pacs: 61-20N, 61-80P
\end{abstract}
\newpage
Most inorganic solids can be made amorphous by vapor deposition onto cold
substrates. However, only a very few of inorganic melts can be 
supercooled by a water or air quench to yield bulk glasses which solidify at
the glass transition temperature $T_g$. Oxides as vitreous silica ($SiO_2$)
and chalcogenides (e.g. $Ge_xSe_{1-x}$) represent some of the best-known glass
formers in nature. 
\par
The origin of glass-forming tendency and the determination of the value of
the glass transition temperature is a subject of great interest, not only
for purely scientific reasons, but also for technological reasons, because
the exact or the approximate value of $T_g$ is needed in some situations 
in order to
optimize the glass preparation process. Therefore, numerous efforts have been
realized in order to understand the nature of glass transition, but also to relate
the value of $T_g$ to some easily measurable quantities. One of the best
known relationships is the "{\em two-third}" rule proposed by Kauzmann, stating that $T_g$
scales with the melting temperature as $T_g\simeq {\frac {2}{3}}T_m$ 
\cite{Kauzmann}. More recently, Tanaka has proposed an empirical relationship
between $T_g$ and the average coordination number. The relationship is 
readily satisfied in oxides, chalcogenides and organic glass-forming
materials \cite{Tanaka}.
\par
Importance of thermodynamic factors in glass formation 
have been discussed by Adam and Gibbs \cite{Adam}, 
and Gibbs and Di Marzio \cite{Gibbs}, suggesting that the glass transition may be
a manifestation of a second-order phase transition, when both the Gibbs
energy and its derivatives remain continous (at the melting temperature, where
the liquid crystallize, the derivatives of the Gibbs energy are discontinuous, 
but not the Gibbs energy). Applying the theory to a liquid
made of molecular chains, the authors found a quantitative relationship 
between the transition temperature and the density of cross-linking agents
inserted inside the system. Later on, the relationship was adapted to
chalcogenide glasses, yielding the following equation, known as the 
Gibbs-Di Marzio law \cite{Sreeram}-\cite{Sreeram1}:
\begin{eqnarray}
\label{1}
T_g\ =\ {\frac {T_0}{1-\beta(<r>-2)}}
\end{eqnarray}
where $T_0$ is the glass transition temperature of the chain-like glass
(e.g. vitreous selenium with $T_0=316\ K$), $\beta$ a system-dependent parameter and 
$<r>$ the average coordination number. The latter has been introduced by
Phillips in his constraint theory \cite{Phillips} and is widely used in 
the investigation of network glasses \cite{Elliott}-\cite{Boolchand}. 
$<r>$ is defined in a M-component glass by 
$<r>=\sum_{i=1}^Mm_ix_i$, where $m_i$ is the valence of an atom with 
concentration $x_i$ (e.g. $<r>=2.67$ in $GeSe_2$).
Agreement of the Gibbs-Di Marzio law with experimental data could be 
obtained by fitting the parameter
$\beta$ for numerous glass systems \cite{Sreeram}-\cite{Sreeram1}. Thus, 
the Gibbs-Di Marzio law seems to describe very well the $T_g$ trends as a 
function of the average coordination number $<r>$, at least for concentrations
corresponding to $<r>\leq 2.7$. 
\par
On the contrary, if one could compute exactly the parameter $\beta$ of 
a given glass system, one could be able to predict its glass transition
temperature $T_g$ as a function of $<r>$. 
The purpose of the present letter is to show that the parameter $\beta$ is related
to the local glass structure and can be easily computed for any chalcogenide
glass system.
\par
In a recent series of papers \cite{Richard}-\cite{Matthieu}, R. Kerner 
and one of us have 
demonstrated by use of a model of statistical agglomeration
that the glass transition temperature $T_g$ displayed a very simple law (called
"{\em slope equations}" in 
weakly modified binary glass systems, such as $Ge_xSe_{1-x}$ with $x$ the concentration
of the modifier atom less than $0.1$. 
\begin{eqnarray}
\label{2}
\biggl[{\frac {dT_g}{dx}}\biggr]_{x=0,T_g=T_0}={\frac {T_0}{\ln \biggl[{\frac
{m_B}{2}}\biggr]}}
\end{eqnarray}
where $T_0$ stands for the initial glass transition temperature and has the 
same meaning as the one appearing in equation (\ref{1}).
$m_B$ is the coordination number (or valence) of the modifier atom 
(e.g. $m_B=4$ is the valence of the germanium atom in $Ge_xSe_{1-x}$ systems).
The value of these coordination numbers can be determined in most of the
situations by the $8-N$ rule,
where $N$ is the number of outer shell electrons of the considered atom
\cite{Mott}.
Thus, the slope in equation (\ref{2}) indicates how the value of the glass 
transition temperature will change if a small proportion of modifier atoms
(such as Ge or As) is added to the initial network. The 
relationship (\ref{2}) is very well satisfied in more than 30 different glass systems
such as chalcogenides and binary glasses (e.g. $(1-x)SiO_2-
xLi_2O$) \cite{EPJB}. Last but not least, if one uses the average coordination 
number of the network $<r>=m_Bx+2(1-x)$ and performs the 
first-order Taylor expansion of the Gibbs-Di Marzio law (\ref{1}) in the
vicinity of $<r>=2$, it is possible to obtain an analytical expression 
for the parameter $\beta$ in two-component glass:
\cite{JNCS97}:
\begin{eqnarray}
\label{3}
{\frac {1}{\beta}}\ =\ (m_B-2)\ln \biggl[{\frac {m_B}{2}}\biggr]
\end{eqnarray}
Here again, the predicted value of the parameter $\beta$ computed from 
obvious structural considerations ($m_B=4$, $m_B=3$, etc.) is in excellent 
agreement with the value determined from experimental data \cite{JNCS97}.
We shall now prove that the factor appearing in
the right-hand side of equation (\ref{3}) has a universal 
character and can be easily extended to a multicomponent chalcogenide 
glass system, 
yielding the value of the parameter $\beta$ for any sytem, to be inserted in 
the Gibbs-Di Marzio law (\ref{1}). To do this, we shall compute from
available experimental data the value of the parameter $\beta$ and compare it
to the predicted one.
\par
Let us extend the expression (\ref{3}) to a M-component glass. For the 
reader's convenience, we shall first consider a glass system made of 
three different kinds of atoms (say $A$, $B$ 
and $C$, with respective concentration $1-x-y$, $x$ and $y$), one of them 
being the atom of the chain-like initial structure (when $x=0$ and $y=0$). The
valences of the involved atoms are $m_A=2$, $m_B$ and $m_C$. The average 
coordination number is $<r>=m_Bx+m_Cy+2(1-x-y)$. We can compute the
derivative with respect to the glass transition temperature:
\begin{eqnarray}
\label{4}
{\frac {d<r>}{dT_g}}\ =\ (m_B-2){\frac {dx}{dT_g}}+(m_C-2){\frac {dy}{dT_g}}
\end{eqnarray}
and look at the limit when $<r>\simeq 2$ (i.e. $T_g\simeq T_0$, $x=0$ and 
$y=0$). Then, we can still identify the derivative of the first-order Taylor 
expansion
of the Gibbs-Di Marzio law (\ref{1}) with the right-hand side of equation 
(\ref{4}), where the quantities $dx/dT_g$ and $dy/dT_g$ have the form
presented in equation (\ref{2}). Identifying all this and simplifying 
by $T_0$ leads to the analytical expression of the parameter $\beta$ in 
a glass made of three components:
\begin{eqnarray}
\label{5}
{\frac {1}{\beta}}=(m_B-2)\ln\biggl[{\frac {m_B}{2}}\biggr]+(m_C-2)\ln \biggl[
{\frac {m_C}{2}}\biggr]
\end{eqnarray}
The extension to multicomponent systems appears to be quite natural. $\beta$
has the same sum rules as the resistance in a parallel circuit in 
electrokinetics, i.e. it is the sum of the $1/\beta$ of each related 
two-component system AB, AC, etc. For a given system made of
$M$ different kinds of atoms with valencies $m_i$, we just have to sum up 
the $M-1$ contributions $(m_i-2)\ln [{\frac {m_i}{2}}]$ in order to obtain
the theoretical value of $\beta^{-1}$.
\begin{eqnarray}
\label{6}
{\frac {1}{\beta}}\ =\ \sum_{i=1}^{M-1} (m_i-2)\ln\biggl[{\frac {m_i}{2}}\biggr]
\end{eqnarray}
In this notation, $m_M$ is the coordination number of the chain atom, equal 
to $2$. We have looked at the validity of this expression on a variety of 
different
glass systems (ternary glasses as $As_xGe_ySe_{1-x-y}$ and multicomponent glasses
such as $As-Si-Ge-Se$, etc.)
\par
{\em Comparison with experimental data:} In order to minimize the influence 
of the preparation techniques and to obtain
a meaningful correlation coefficient, we have carefully selected
data of i) glass systems prepared with the same heating/cooling rate ii) glass
systems with more than five different compositions. 
We have checked that the influence of the heating rate on the value of $\beta$
could be neglected at the heating rates which were used in the preparation of the
glasses we have investigated. To do this, we have used both the Kissinger's 
formula \cite{Kissinger} 
\begin{eqnarray}
\label{7}
\ln\biggl[{\frac {T_g^2}{Q}}\biggr]+const.&=&{\frac {E}{RT_g}}
\end{eqnarray}
which is very well adapted for the description of 
chalcogenide glasses, and the Gibbs-Di Marzio law.
$Q$ is the heating rate, $E$ the activation energy for glass transition
and $R$ the gas constant. Inserting typical values of $<r>$, $E$ and $T_0$
shows that $\beta$ depends weakly on $Q$ ($\Delta\beta/\beta\leq 5\%$).
This verification has been also realized numerically on systems
for which the heating rate in DSC calorimetry was reported. For example, 
the parameter $\beta$ of the glass $As-Sb-Se$ lies in 
the range [1.14, 1.17] (computed from a least-squares fit) for 
cooling rates between $Q=0.62\ K.min^{-1}$ and $40\ K.min^{-1}$, which corresponds
to $\Delta \beta /\beta =0.013$.
\par
The initial value $T_0$
has been averaged over a set of data found in the literature ($T_0$ of
$v-Se$ has been taken as $316\ K$, of $v-S$ as $245\ K$). We have performed a 
least-squares fit of the Gibbs-Di Marzio law applied to the data. The results
of the fit for the parameter $\beta$, denoted as $\beta_{exp}$, and the
correlation coefficient are displayed in Table I with their corresponding 
reference. For completeness, we have reported in the first part of Table I some
results \cite{JNCS97} for two-component systems, which satisfy (\ref{3}).
The results can be compared with the predicted value of $\beta$, denoted
as $\beta_{pr}$ in Table I, which has been computed from equation (\ref{6}).
For example, in the system Si-As-Ge-Te the involved valencies of the modifier 
atoms are 4, 3 and 4 respectively. Thus, $\beta_{pr}^{-1}=3\ln 2+\ln 3$ 
and $\beta_{pr}=0.31$, in excellent agreement with the fit $\beta_{exp}=0.30$.
The number of different glasses satisfying exactly or roughly the rule
(\ref{6}) is impressive and proves that the agreement is not a matter of
coincidence.
\par
Let us now come to the prediction of glass transition temperature values. We can
note that the addition of a two-valenced atom (as tellurium or sulphur) in 
multicomponent glass systems does not affect the value of the parameter $\beta$.
A 4-component system which involves a two-valenced atoms can therefore be
considered as a ternary system (Table I) and $T_0$ is then the glass 
transition temperature of the initial mixture (e.g. $Se-Te$).
But for all the other elements of the columns III, IV and V, we can compute the 
value of $\beta$ and represent $T_g$ as a function of $<r>$ (fig.1). The 
simultaneous use of equations (\ref{1}) and (\ref{6}) should give the value
of the glass transition temperature of any composition, at least for 
$<r>\leq 2.4$ (as shown on figure 1 for the comparison between the 
theoretical Gibbs-Di Marzio law and the experimental data). For greater
values of $<r>$, one should take into account intermediate range order 
effects such as the existence of rings \cite{Angell}-\cite{Boolchand}. This
extra influence has not been considered here.
\par
{\em Influence of a demixed structure:} Nevertheless, there are still
several systems for which the predicted value of $\beta$ does not match 
with the computed one. The difference between $\beta_{exp}$ and $\beta_{pr}$
can be quantitatively discussed in terms of the presence of some demixed 
structure inside the network. This feature is generally
used in order to explain the unusual variation of $T_g$ in glasses, such as 
$Bi_xSe_{1-x}$ or $In_xSe_{1-x}$. In the former system, data obtained for
the $Bi-Se$ glass, show that glass transition temperature increases with
$Bi$ content for low modification, in perfect agreement with the slope 
equation (\ref{2}). But when $x>0.1$, $T_g$ remains constant
\cite{Myers}. This behavior is attributed to the existence of a demixed
structure of $Bi_2Se_3$ microclusters. The same happens in the indium selenide
compound at every concentration, and evidence of $In_2Se_3$ clusters has 
also been discussed \cite{Saiter}. 
\par
Again, let us consider as before a ternary glass system $A_{1-x-y}B_xC_y$ 
with the corresponding coordination numbers $m_A=2$, $m_B$ and $m_C$.
If we assume that the general tendency of the glass is to form a demixed
strcuture of $A$ and $B$ in stoichiometric proportions, then we can rewrite
the system as: $(B_{m_A}A_{m_B})_{x/m_A}C_yA_{1-y-(m_A+m_B)x/m_A}$ which
defines an effective concentration of $C$ atoms $y^{eff}={\frac {y}{1-x(m_A+
m_B)/m_A}}$
and the average coordination number of the system is given by:
\begin{eqnarray}
\label{8}
<r>&=&m_A+{\frac {(m_C-m_A)y}{1-x{\frac {(m_A+m_B)}{m_A}}}}
\end{eqnarray}
If we proceed as before, i.e. taking the derivative with respect to $T_g$ and
looking at the limit ($x,y\rightarrow 0$) in order to identify with the 
expansion of the Gibbs-Di Marzio law, we can see that the corresponding
parameter $\beta$ is defined as $\beta^{-1}=(m_C-m_A)\ln [{\frac {m_C}{m_A}}]$.
In other words, the parameter $\beta$ of a ternary system which displays a 
demixed structure can be computed by considering only the remaining two-component
glass. Here again, we have checked the validity of this rule on a set of
germanium incorporated chalcogenides (Table II). We have considered all
possible stoichiometric demixed structures at the tie-line composition
(e.g. in $Ge_xAs_ySe_{1-x-y}$, the possible structures are $GeSe_2$ and
$As_2Se_3$). Again, we have computed the corresponding parameter
$\beta_{pr}$ and compared it to the value $\beta_{exp}$ obtained from
a least-squares fit. Table II shows that most of the $III-IV-VI$ systems
such as $Sb-Ge-S$ glasses behave as a single $IV-VI$ glass (as $Ge-Se$) with 
parameter $\beta_{pr}=0.72$. This is explained by the presence of $A_2X_3$
($X=S,Se,Te$) clusters with $A$ an element of the column III or V. 
\par
In summary, we would like to recall the most important rules to be used if one 
wishes to determine a glass transition temperature of a multicomponent
chalcogenide glass.
\begin{enumerate}
\item Consider all the possible demixed stoichiometric structures of the
M-component chalcogenide glass. There is in general a strong evidence of
demixing in the case of $Sb$ ternary based glasses \cite{Sb2S3}.
\item Given the concentrations $x_i$ of the atoms, compute the average 
coordination number $<r>$, either directly, or by use of the effective 
concentration $x_i^{eff}$ if there is some evidence of structural
demixing.
\item Compute the parameter $\beta$ from the coordination numbers of the
remaining atoms (those which are not involved in the demixed structure).
\item Insert $\beta$ and $<r>$ in the Gibbs-Di Marzio law in order to obtain
the glass transition temperature.
\end{enumerate}
As illustrative examples of these rules, we give for the conclusion glass
transition temperatures of systems, which have to our knowledge never been
investigated.\par 
Assuming that $\Delta \beta /\beta \simeq 0.1$ (possibly produced by the 
heating rate effects, see above, or anything else) and if there is no 
demixing, the glass $As_5Ga_5Si_{15}Se_{75}$ 
should have a transition temperature of about $385\pm 8\ K$ since $\beta=0.45$ is computed
from the valencies $m_i=(3,3,4)$ and $<r>=2.4$. In the sulphide analog glass, 
one should measure $299\pm 7\ K$. Similarly,
the glass $Ga_{10}B_{10}S_{80}$ should have a transition temperature
of $T_g=325\pm 11\ K$ ($\beta=1.23$ computed from the valencies $m_i=(3,3)$ 
and $<r>=2.2$).
\par
The influence of intermediate range order on this set of rules will be 
discussed in a forthcoming article. Also, work devoted to the influence of 
the cooling/heating rate on the value of the glass transition temperature is 
in progress.
\par
The authors gratefully acknowledge R. Kerner, R.A. Barrio and J. Ledru for 
stimulating discussions. This work has been supported by the CONACyT grant 
No. 25237-E.

\newpage
\begin{table}
\begin{center}
\begin{tabular}{c|ccccc}
System&$\beta_{pr}$& &$\beta_{exp}$&Correlation&Reference \\
 & & & &coefficient& \\ \hline\hline
Ge-(Se)&0.72& &0.72&0.988&from \cite{JNCS97}\\
Ge-(S)&0.72& &0.73&0.998&from \cite{JNCS97}\\
Si-(Se)&0.72& &0.81&0.997&from \cite{JNCS97}\\ \hline
Ge-Sb-(Se)&0.56& &0.66&0.972&\cite{Sreeram}\\
Ga-Ge-(Se)&0.56& &0.55&0.965&\cite{Giridhar}\\
Ga-Ge-(S)&0.56& &0.59&0.823&\cite{Saffarini}\\
As-Sb-(Se)&1.23& &1.17&0.995&\cite{Mahadevan} \\
Al-P-(Se)&0.32& &0.21&0.952&\cite{Hudalla}\\
Ge-Sb-Te-(Se)$^*$&0.56& &0.55&0.998&\cite{Sreeram}\\
Si-As-Ge-(Te)&0.31& &0.30&0.979&\cite{ElFouly}\\
Ge-Sb-As-Te-(Se)$^*$&0.45& &0.55&0.989&\cite{Sreeram}\\
\end{tabular}
\vspace{0.5cm}
\caption{Different multicomponent glass systems. Comparison between the
predicted value $\beta_{pr}$ obtained from equation (\ref{6}) given the 
valencies of the involved atoms, and the value $\beta_{exp}$ computed from
experimental data by a least-squares fit. The atom which corresponds to the
chain-like entity corresponds to the chemical symbol inside the bracketts.
M-component glasses with an $^*$ can be considered as (M-1) component glasses
for the computation of $\beta$, since they involve a two-valenced atom
(Te).}
\end{center}
\end{table}
\begin{table}
\begin{center}
\begin{tabular}{c|cccc}
System&$\beta_{exp}$&Demixed&Correlation&Reference \\
& &structure&coefficient& \\ \hline\hline
In-Ge-(Se)&0.77&$In_2Se_3$&0.993&\cite{Giridhar}\\
Sb-Ge-(S)&0.61&$Sb_2S_3$&0.932&\cite{ElHamalawy}\\
Sb-Ge-(Se)&0.78&$Sb_2Se_3$&0.986&\cite{Fouad}\\
Sb-Ge-(Te)&0.79&$Sb_2Te_3$&0.992&\cite{Lebaudy}\\
Sn-Ge-(Se)&0.68&$SnSe_2$&0.973&\cite{Haruvi}\\
\end{tabular}
\vspace{0.5cm}
\caption{Different multicomponent glass systems exhibiting a demixed structure. 
Comparison between the predicted value of $\beta_{pr}=0.72$ obtained by 
considering the remaining $Ge-X$ system ($X=S,Se,Te$)
and the value $\beta_{exp}$ computed from experimental data by a 
least-squares fit. The atom which corresponds to the
chain-like entity corresponds to the chemical symbol inside the bracketts.} 
\end{center}
\end{table}

\newpage
\begin{figure}
\begin{center}
\psfig{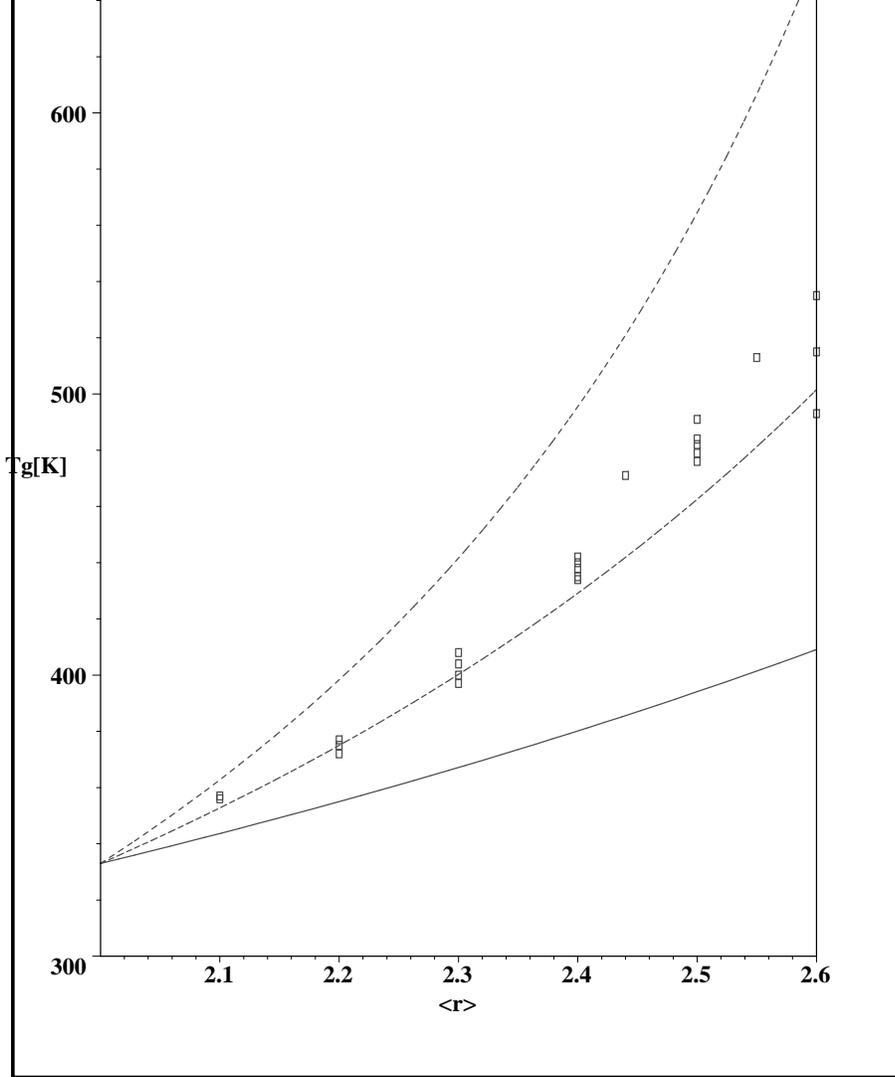}
\caption{Several Gibbs-Di Marzio laws with the computed value of $\beta$
in multicomponent chalcogenide glasses. Solid line: $m_i=(3,4,4)$ and $\beta=
0.31$. Dashed line: $m_i=(2,3,4)$ and $\beta=0.56$ with the experimental 
data of the $Te-Sb-Ge-Se$ system [3] (boxes). Dotted line: $m_i=(3,3,3)$ and 
$\beta=0.82$.}
\end{center}
\end{figure}


\begin{thebibliography}{100}
\bibitem{Kauzmann} W. Kauzmann, Chem. Rev. {\bf 43} (1948) 219
\bibitem{Tanaka} S. Tanaka, Solid State Comm. {\bf 54} (1984) 867
\bibitem{Adam} G. Adam and J.H. Gibbs, J. Chem. Phys. {\bf 43} (1965) 139
\bibitem{Gibbs} J.H. Gibbs and E.A. Di Marzio, J. Chem. Phys. {\bf 28} (1958) 373
\bibitem{Sreeram} A.N. Sreeram, D.R. Swiler, A.K. Varshneya, J. Non-Cryst. 
Solids {\bf 127} (1991) 287
\bibitem{Sreeram1} U. Senapati and A.K. Varshneya, J. Non-Cryst. Solids {\bf 197}
(1996) 210
\bibitem{Phillips} J.C. Phillips, Phys. Rev. B{\bf 31} (1985) 8157
\bibitem{Elliott} S.R. Elliott, {\em The physics of amorphous materials}, Wiley (NY)
1989
\bibitem{Angell} M. Tatsumisago, B.L. Halfpap, J.L. Green, SM. Lindsay
and C.A. Angell, Phys. Rev. Lett. {\bf 64} (1990) 1549
\bibitem{Boolchand} W. Bresser, P. Boolchand and P. Suranyi, Phys. Rev. Lett.
{\bf 56} (1986) 2493; P. Boolchand, Phys. Rev. Letters {\bf 57} (1986) 3233;
P. Boolchand and J.C. Phillips, Phys. Rev. Letters {\bf 68} (1992) 252
\bibitem{Richard} R. Kerner, Physica B {\bf 215} (1995) 267
\bibitem{JNCS97} R. Kerner and M. Micoulaut, J. Non-Cryst. Solids {\bf 210}
(1997) 298
\bibitem{EPJB} M. Micoulaut, Eur. Phys. J. B {\bf 1} (1998) 277
\bibitem{Matthieu} M. Micoulaut, J. Non-Cryst. Solids (1998) in press
\bibitem{Mott} N.F. Mott, Philos. Mag. {\bf 19} (1969) 835
\bibitem{Kissinger} H.E. Kissinger, J. Res. Nat. Bur. Stand. {\bf 57} (1956) 217
\bibitem{Giridhar} A. Giridhar and S. Mahadevan, J. Non-Cryst. Solids {\bf 151} 
(1992) 245
\bibitem{Saffarini} G. Saffarini, Solid State Comm. {\bf 91} (1994) 577
\bibitem{Mahadevan} A. Giridhar and S. Mahadevan, J. Non-Cryst. Solids {\bf 51} 
(1982) 305
\bibitem{Hudalla} C. Hudalla, B. Weber and H. Eckert, J. Non-Cryst. Solids
{\bf 224} (1998) 69
\bibitem{ElFouly} M.H. El-Fouly, A.F. Maged, H.H Amer and M.A. Morsy,
J. Mater. Sci. {\bf 25} (1990) 2264
\bibitem{Myers} M.B. Myers, J.C. Schotmiller and W.J. Hillegas, Anal. Calor.
{\bf 2} (1970) 309
\bibitem{Saiter} J.M. Saiter, J. Ledru, G. Saffarini, S. Benazeth, Materials
Letters {\bf 28} (1996) 451
\bibitem{Giridhar} A. Giridhar and S. Mahadevan, J. Non-Cryst. Solids {\bf 151}
(1992) 245
\bibitem{ElHamalawy} A.A. El-Hamalawy, M.M El-Zaidia, A.A. Amman and M. Elkhdy, J. Mater.
Sci. Mater. Electron. {\bf 5} (1994) 147
\bibitem{Fouad} S.S. Fouad, J. Phys. D Appl. Phys. {\bf 28} (1995) 2318
\bibitem{Lebaudy} P. Lebaudy, J.M. Saiter, J. Grenet, M. Belhadji and C. 
Vautier, Mater. Sci. Eng. A{\bf 132} (1991) 273
\bibitem{Haruvi} I. Haruvi-Busnach, J. Dror and N. Croitoru, J. Mater. Sci.
{\bf 5} (1990) 1215
\bibitem{Sb2S3} Y. Sagara, O. Uemura, Y. Suzuki and T. Satow, Phys. Stat. 
Solidi A{\bf 33} (1976) 691
\end{thebibliography}
\end{document}